\documentclass[A4paper,12pt]{article}
\usepackage[a4paper,total={170mm,257mm},left=20mm,top=20mm,]{geometry}
\usepackage{amsmath,amsthm,amssymb,mathtools}
\usepackage{natbib}
\usepackage[colorlinks,citecolor=blue,urlcolor=blue]{hyperref}
\usepackage{subcaption} 
\usepackage{xcolor}

\newtheorem{theorem}{Theorem}

\usepackage{tikz}
\usetikzlibrary{arrows,automata,er,positioning,calc,arrows.meta,shapes,decorations,fit}
\tikzset{
    -Latex,auto,node distance =1 cm and 1 cm,semithick,
    state/.style ={ellipse, draw, minimum width = 0.7 cm},
    point/.style = {circle, draw, inner sep=0.04cm,fill,node contents={}},
    bidirected/.style={Latex-Latex,dashed},
    el/.style = {inner sep=2pt, align=left, sloped}
}

\begin{document}

\def\T{\mbox{\scriptsize \rm T}}
\def\E{\mbox{\rm E}}
\def\Var{\mbox{\rm Var}}
\def\Cov{\mbox{\rm Cov}}
\def\Corr{\mbox{\rm Corr}}
\def\Pr{\mbox{\rm Pr}}
\def\I{\mbox{\rm I}}
\def\bbeta{\mbox{\boldmath${\beta}$}}
\def\balpha{\mbox{\boldmath${\alpha}$}}
\def\bgamma{\mbox{\boldmath${\gamma}$}}
\def\sbeta{\mbox{\scriptsize \boldmath${\beta}$}}
\def\eeps{\mbox{\boldmath${\epsilon}$}}
\def\bmu{\mbox{\boldmath${\mu}$}}
\def\bxi{\mbox{\boldmath${\xi}$}}
\newcommand{\bgams}{\mbox{\boldmath{\scriptsize $\gamma$}}}
\newcommand{\bLam}{\mbox{\boldmath{$\Lambda$}}}
\newcommand{\Ell}{\mbox{$\cal{L}$}}
\newcommand{\cG}{\mathcal{G}}
\newcommand{\cL}{\mathcal{L}}
\newcommand{\half}{\frac{1}{2}}
\newcommand{\bsh}{\parindent 0em}
\newcommand{\esh}{\parindent 2.0em}
\newcommand{\eps}{\epsilon}
\newcommand{\peps}{{(\eps)}}
\newcommand{\bzro}{{\bf 0}}
\def\bU{{\bf U}}
\def\bV{{\bf V}}
\def\bG{{\bf G}}
\def\bC{{\bf C}}
\def\bD{{\bf D}}
\def\bu{{\bf u}}
\def\bx{{\bf x}}
\newcommand{\cA}{\mathcal{A}}
\newcommand{\cF}{\mathcal{F}}
\newcommand{\cW}{\mathcal{W}}
\newcommand{\cH}{\mathcal{H}}
\newcommand{\cI}{\mathcal{I}}
\newcommand{\cD}{\mathcal{D}}
\newcommand{\cC}{\mathcal{C}}
\newcommand{\fC}{\mathfrak{C}}
\newcommand{\fL}{\mathfrak{L}}
\newcommand{\cV}{\mathcal{V}}
\newcommand{\cM}{\mathcal{M}}
\newcommand{\cE}{\mathcal{E}}
\newcommand{\cS}{\mathcal{S}}
\newcommand{\cR}{\mathcal{R}}
\newcommand{\cQ}{\mathcal{Q}}
\newcommand{\cB}{\mathcal{B}}
\newcommand{\cP}{\mathcal{P}}
\newcommand{\cY}{\mathcal{Y}}
\newcommand{\cN}{\mathcal{N}}
\newcommand{\PP}{\mathbb{P}}
\newcommand{\Pn}{\mathbb{P}_n}
\newcommand{\msp}{\mathcal{P}}
\newcommand{\Kh}[1]{K\left(\frac{#1}{h}\right)}
\newcommand{\KhX}{\Kh{X_{Pi}-s}}
\newcommand{\sd}{\, \begin{picture}(-1,1)(-1,-2)\circle*{2}\end{picture}\ }
\newcommand{\sumi}{\sum_{i=1}^n}
\newcommand{\sumsti}{\sum_i^*}
\newcommand{\sumj}{\sum_{j=1}^n}
\newcommand{\fmh}{\frac{1}{mh}}
\newcommand{\pr}{\prime}
\newcommand{\frsm}[2]{\mbox{\small $\frac{#1}{#2}$}}
\newcommand{\frsms}[2]{\mbox{\scriptsize $\frac{#1}{#2}$}}
\newcommand{\real}{\mathbb{R}}
\newcommand{\cp}{\stackrel{p}{\rightarrow}}
\newcommand{\rta}{\rightarrow}
\newcommand{\omitt}[1]{{}}
\newcommand{\pss}[1]{{(#1)}}
\newcommand{\uhi}{\hat{U}_i}
\newcommand{\uhm}{\bar{\hat{U}}}
\newcommand{\vpp}{\vspace*{6pt}}
\newcommand{\tv}{\tilde{v}}
\newcommand{\didl}{\vphantom{\bigodot}}

{
	\title{\bf Cumulative Incidence Function Estimation \\ Based on Population-Based Biobank Data}
	\author{Malka Gorfine\\
		Department of Statistics and Operations Research\\
		 Tel Aviv University, Israel\\
\vspace{0.5cm} \\  
    David M. Zucker \\
    Department of Statistics and Data Science\\
    The Hebrew University of Jerusalem, Israel\\
\vspace{0.5cm}   \\ 
    Shoval Shoham\\
    Department of Statistics and Operations Research\\
    Tel Aviv University, Israel.
    }
	\date{}	
	\maketitle
}

\abstract{
Many countries have established population-based biobanks, which are being used increasingly in epidemiolgical and clinical research. These biobanks offer
opportunities for large-scale studies addressing questions beyond the scope of traditional clinical trials or cohort studies. However, using biobank data poses new challenges. Typically, biobank data is collected from a study cohort recruited over a defined calendar period, with subjects entering the study at various ages falling between $c_L$ and $c_U$. 
This work focuses on biobank data with individuals reporting disease-onset age upon recruitment, termed prevalent data, along with individuals initially recruited as healthy, and their disease onset observed during the follow-up period. We propose a novel cumulative incidence function (CIF) estimator that efficiently incorporates prevalent cases, in contrast to existing methods, providing two advantages: (1) increased efficiency, and (2) CIF estimation for ages before the lower limit, $c_L$. }

{\bf keywords:} 
Aalen-Johansen estimator, Delayed entry, Illness-death model, Left truncation, Survival analysis. 

\newpage

\section{Introduction}
\begin{figure}
	\begin{center}
		\begin{tikzpicture}
		\node[state, ellipse] (1) at (0,0) {(0) Healthy};
		\node[state, ellipse] (2) [right = of 1] {(1) Disease};
		\node[state, ellipse] (4) [below = of 2] {(2) Death};		
		\path (2) edge (4) ;
		\path (1) edge  (2);
		\path (1) edge (4);
		\end{tikzpicture}
	\end{center}
	\caption{{Semi-competing risks of a chronic disease}}\label{Fig:multis}
\end{figure}
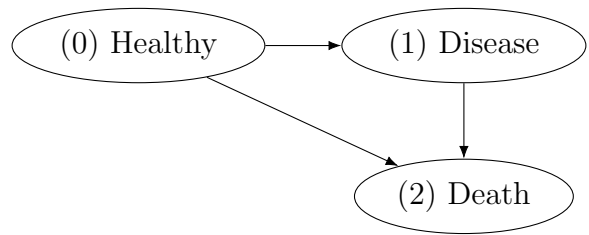

We consider the problem of estimating the incidence of some disease using  population-based biobank data.
We work with  the illness-death model of Fig. \ref{Fig:multis}, also known as the
semi-competing risks model.
Here, individuals start in a healthy state and can then move to the diseased state, signifying being diagnosed with the disease of interest, and from there to the dead state. Individuals can also move directly from the healthy state to the dead state. Each arrow in the figure represents a different hazard function. Let the random variables
$T_1$ and $T_2$ represent the age at diagnosis and age at death, respectively. Assume for simplicity that $(T_1,T_2)$ are  absolutely continuous. Since the
disease cannot occur after death, the density function of $(T_1, T_2)$ is concentrated on 
the upper wedge $t_2 \geq t_1$. The joint density of $(T_1, T_2)$, denoted by
$f_{T_1,T_2}(t_1, t_2)$, is defined for $t_2 \geq t_1 \geq 0$ and
$$
\int_0^{\infty} \int_{t_1}^{\infty} f_{T_1,T_2}(t_1, t_2) dt_2 dt_1 = \Pr (T_1 < \infty) \leq 1 \, .
$$
For those who died without having the disease, we set $T_1 = \infty$, and the conditional density of $T_2$ is defined over $t_2 > 0$. This description of the model specifies no
probability content in the lower wedge $t_2 < t_1 < \infty$, which is a true reflection of
the physical situation, and is therefore recommended for semi-competing risks, as
described by \cite{xu2010statistical}.

Our current goal is to use biobank data to estimate the probability of having the disease by time $t$, which is the cumulative incidence function (CIF), 
given by
$$
G_1(t) = \Pr (T_1 \leq t, T_2 > T_1) \, \,\,\, t \in [0,\tau]
$$
for a constant $\tau >0$ representing the maximum age at end of follow-up,
or alternatively, the conditional CIF,
$$
G_1(t|T_2>c_L) = \Pr (T_1 \leq t, T_2 > T_1|T_2>c_L) \, \,\,\, t \in [0,\tau]
$$
for a constant $c_L>0$ representing the minimum age of entry into the study.

Population-wide biobanks have been established in various countries, such as the UK, Sweden, Denmark, Canada, South Korea, China, Japan, Singapore, and the USA. These extensive repositories  gather, analyze, and store phenotypic and genetic information from representative samples of their respective populations. Medical data recorded routinely in these biobanks are increasingly employed for research purposes. Their utilization presents an opportunity for large-scale, high-quality studies that can address questions not easily tackled by randomized clinical trials or classical cohort studies involving bespoke data collection. However, the use of biobank data also introduces new challenges. The aim of this work is to solve one open problem related to analysis of time-to-event data, that represents a major leap forward in the area of survival analysis in general, and in analyzing biobank data in particular.

Typically, biobank data are collected from a study cohort recruited over a defined calendar period, with subjects entering the study at various ages falling between specified limits $c_L$ and $c_U$. In the UK Biobank (UKB) data, for instance, $c_L$ is set at 40, and $c_U$ at 69. Following recruitment, the biobank participants undergo prospective follow-up. This study design introduces certain ascertainment-related challenges. Firstly, there is no information available regarding deaths occurring before the minimal recruitment age, $c_L$. Secondly, the data represents only the segment of the population that survived up to the recruitment age, a challenge recognized as left truncation or delayed entry. Addressing these issues is crucial to prevent biased results, see \cite{keiding1991age,saarela2009joint,vakulenko2016comparing} among others.

When UKB data are used to estimate the conditional CIF, $G_1(t|T_2>c_L)$, of a specific disease, participants are categorized into three groups: ``prevalent" cases—individuals already diagnosed with the disease at the time of recruitment; ``incident" cases—those diagnosed during the follow-up period; and right-censored observations—individuals who were not diagnosed with the disease by the conclusion of the follow-up (or analysis time). For instance, among the current 502,420 observations in the UKB, there are a total of 383 cases of acute myeloid leukemia (AML) cancer, with 20\% of them classified as prevalent.

Estimating from left-truncated and right-censored data is typically done either through risk-set correction or inverse-probability weighting (IPW). The well-known Aalen-Johansen estimator \citep{aalen1978empirical} can be easily adjusted for delayed entry (see \cite{allignol2010note} and Section 2.1 below), where the adjustment is based on risk-set correction. An individual enters the risk set after its left-truncation (i.e., enrollment) time, and the individual stays in the risk set until its event or censoring event time, whichever comes first. However, this risk-set adjustment omits prevalent cases from the estimation procedure, since the event time precedes the entry time. Besides the potential reduction in efficiency, it also implies that the Aalen-Johansen estimator can estimate CIF only for $t > c_L$, conditional on being event-free at the time of recruitment. 

\begin{figure}
\centering
\includegraphics[width=0.5\textwidth]{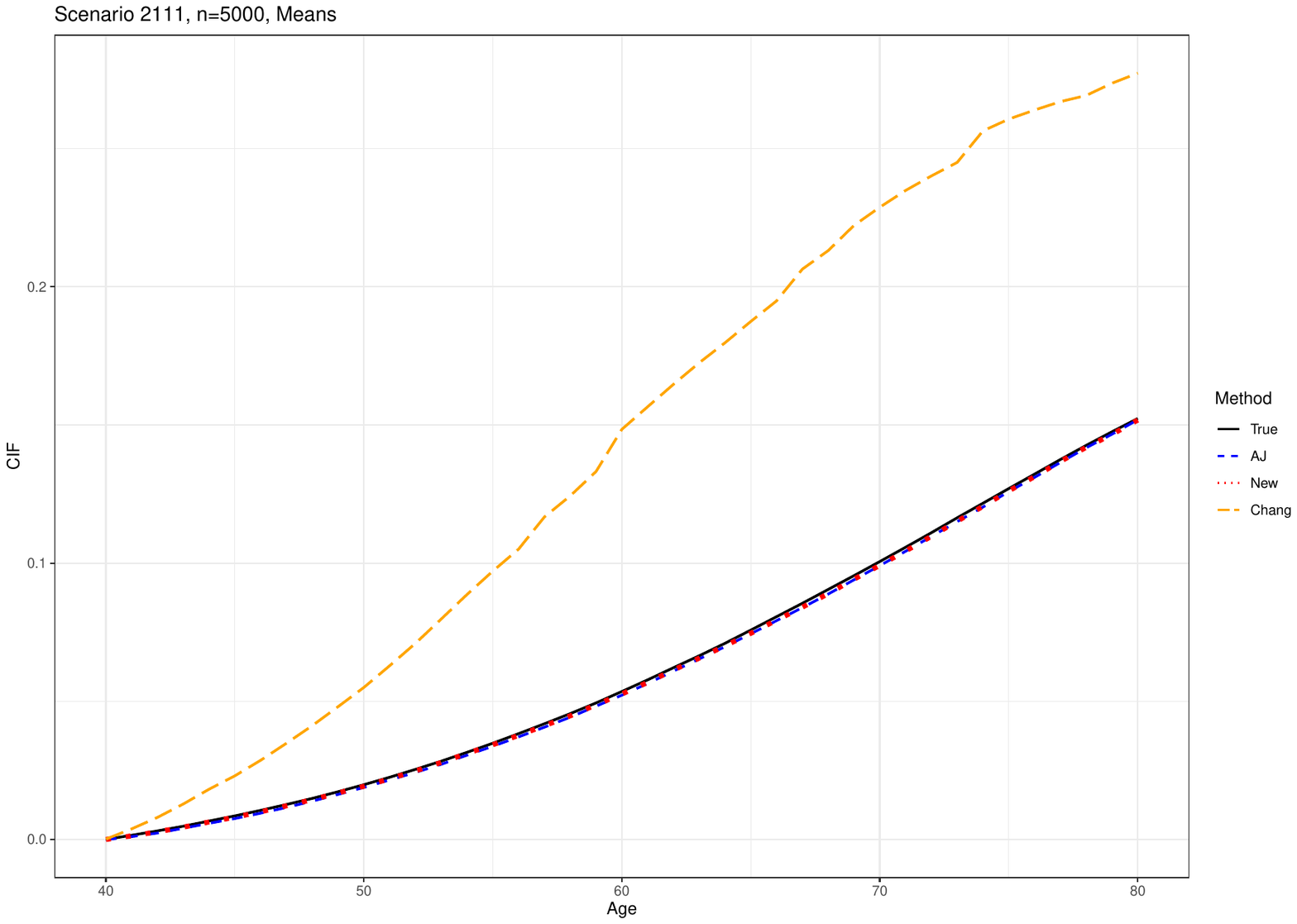}
\caption{Simulation results of configuration 2111 demonstrating a substantial biased results of \cite{chang2006nonparametric} (Chang) versus Aalen-Johansen (AJ) and the proposed estimators (New). \label{fig:exampleChang} }
\end{figure}

\cite{chang2006nonparametric} and \cite{vakulenko2017nonparametric} provided nonparametric estimators employing IPW methods. Unlike the Aalen-Johansen estimator, these approaches do not exclude prevalent observations, but correct for
the sampling bias by IPW. Their methods assign positive
masses only to completely uncensored observations, and the weights
depend on the distributions of censoring and truncation.  However, both methods require the condition that the upper limit of the support of the distribution of age at censoring be  greater than the upper limit of the support of the distribution of age at death. If this condition is violated, the estimators can be very seriously biased. As an example, Fig. \ref{fig:exampleChang} presents a graph of the mean of the Chang-Zheng estimator along with the true CIF, the mean of the Aalen-Johansen estimator, and the mean of our proposed estimator for the setting of Scenario 2111 in our simulation study below. The bias of the Chang-Zheng estimator is pronounced. As a result, we will not explore these works further.

Our work here presents a solution to the problem of how to use the information from the prevalent cases without imposing the foregoing restrictive condition on the support of the age at censoring. We introduce a novel CIF estimator that efficiently utilizes the entire available data, including prevalent cases. The inclusion of prevalent cases offers two advantages: (1) increasing the number of observed events, and (2) estimating CIF at ages prior to age $c_L$. This can be particularly valuable for diseases with low fatality rates at early ages.

\section{CIF Estimators}
Consider $n$ independent observations. Let $T_{1i}$ and $T_{2i}$ be the age at diagnosis and
age at death, respectively, of the $i$th observation, $i = 1,\ldots,n$. 
If the participant dies without being diagnosed with the disease under study, we set $T_{1i} = \infty$. Define $C_i$ as the right-censoring time and $R_i$ as the age at recruitment, where $c_L \leq R_i \leq c_U$ signifying that subjects are recruited at ages between $c_L$ and $c_U$. We focus on the homogeneous case, i.e., without covariates. We assume that $C_i$ is independent of $(T_{1i}, T_{2i})$ and that $R_i$ is independent of the age of disease onset and quasi-independent \citep{tsai1990testing} of death and censoring ages. Define $V_{1i} = \min(T_{1i}, T_{2i}, C_i)$, $V_{2i}=\min(T_{2i},C_i)$, $\delta_{1i} = I(T_{1i} \leq \min(T_{2i}, C_{i}))$,
and $\delta_{2i} = I(T_{2i} \leq C_{i})$. The observed data consist of $\{V_{1i},V_{2i}, \delta_{1i}, \delta_{2i}, R_i \, , \, i=1,\ldots,n\}$. 
 
\subsection{Left-Truncation Adjusted Aalen-Johansen Estimator}
Denote the hazard functions of transitions $0 \rightarrow k$, $k=1,2$ (see Fig. \ref{Fig:multis}) by
$$
\lambda_{k}(t) = \lim_{dt \downarrow 0} (dt)^{-1} \Pr(t \leq T_{k} < t+dt| T_1 \geq t, T_2 \geq t) \, .
$$
$G_1(t)$ can then be written as
$$
G_1(t) = \int_0^t \lambda_{1}(u) \Pr(T_1 \geq u, T_2 \geq u) du = \int_0^t \lambda_{1}(u) \Pr(T^* \geq u) du \, ,
$$
where $T^*=T_1 \wedge T_2$. 
Let $N_{1i}(u) = \delta_{1i} I(V_{1i} \leq u)$ be the counting process for disease occurrence, and let $Y_{1i}(u)$ be the at-risk indicator, equal to 1 if the particicipant is still at risk and equal to 0 if not.
The Aalen-Johansen estimator of the CIF is then given by
$$
\widehat{G}^{AJ}_1(t)= \int_0^t \hat{S}_{T^*}(u-) \frac{\sumi dN_{1i}(u)}{\sumi Y_{1i}(u)} \, ,
$$
where $\widehat{S}_{T^*}(\cdot)$ is the Kaplan-Meier estimator of the survival function of $T^*$. For data with left censoring, estimation is based on risk-set adjustment for delayed entry. Namely, at time $t$, the at-risk process of individual $i$ is defined by $Y_{1i}(t) =I(R_i \leq t \leq V_{1i})$. Clearly, this estimator
excludes prevalent cases, i.e., individuals with $V_{1i} < R_i$. Therefore, the estimand of $\widehat{G}^{AJ}_1(t)$ is the CIF given that $T_1 \geq c_L$ and $T_2 \geq c_L$, namely, $G_1(t|T_1 \geq c_L, T_2 \geq c_L)$. The estimator $\widehat{G}^{AJ}_1(t)$ is a consistent estimator of this estimand.

\subsection{The Proposed Estimator}
We redefine the CIF function as $G_1(t)=\Pr(T_1 \leq t_1, T_2 > T_1, T_2 \leq \tau)$ and assume that the probability of surviving up to time $\tau$ is positive but small, so that the bias introducedby including the event $T_2 \leq \tau$ is minimal. Then, one can write  
\begin{align}
G_1(t_1) & = \int_0^{\tau} \lim_{dt \downarrow 0} (dt)^{-1} 
\Pr (T_1 \leq t_1 \wedge t_2, T_2 \in [t_2,t_2+dt)) dt_2 \nonumber \\
& = \int_0^{\tau} F_{1|2}( t_1 \wedge t_2 |t_2)  dF_{2}(t_2) 
\label{gee}
\end{align}
where $F_{1|2}(t_1|t_2)=1-S_{1|2}(t_1|t_2)=1-\Pr(T_1 > t_1| T_2=t_2)$ and $F_{2}(\cdot)$ is the cumulative distribution function of age at death. By focusing on the conditional distribution of age at diagnosis, given age at death, we can easily accommodate delayed entry, as explained in further detail below.

It is easy to verify that since $R_i$ is independent of $T_{1i}$ and quasi independent of $T_{2i}$, given the death age, the recruitment age has no additional predictive value for the age at onset $T_{1i}$. Specifically,   for any $r \in [c_L,c_U]$, $t_2 > t_1 \geq 0$ and  $ t_2 > r$, 
$$
S_{1|2}(t_1|t_2) = \Pr(T_1 > t_1 | T_2=t_2, R=r)   \, .
$$
Let the disease-at-risk process, adjusted for delayed entry, be defined by
$$
Y_i(t_1,t_2) = I(t_1 \leq V_{1i})I(R_i \leq t_2) \hspace{0.5cm} i=1,\ldots,n
$$
such that subject $i$ is at risk at $(t_1,t_2)$ if the subject is alive, non-censored and free of the disease by time $t_1$ and is also recruited by time $t_2$. The disease counting process is again defined by $N_{1i}(t_1)=\delta_{1i}I(V_{1i} \leq t_1)$.

We now need to construct estimators of $F_{1|2}(t_1|t_2)$ and $F_2$. In regard to
$F_2$, we take
$\widehat{F}_2(t)=1-\widehat{S}_2(t)$ and $\widehat{S}_2(\cdot)$ is the Kaplan-Meier (KM) estimator of death based on $\{V_{2i},\delta_{2i},R_i \, , \, i=1,\ldots,n \}$ with the risk-set correction for left-truncated and right-censored data. In regard to $F_{1|2}$, we proceed
as follows. Given $T_{2i}$, by standard martingale theory one can write
\begin{equation}\label{eq:mart}
dM_{i}(t_1) = dN_{1i}(t_1) - Y_i(t_1,T_{2i}) \Lambda_{1|2}(dt_1|T_{2i}) \hspace{0.5cm} t_1 < T_{2i}     
\end{equation}
where $\Lambda_{1|2}(t_1|t_2) = -\log S_{1|2}(t_1|t_2)$ and $M_i(t_1)$ is a zero-mean martingale with respect to the filtration 
$$
\mathcal{F}^{(i)}_{t_1,t_2} = \{R_i,T_{2i},N_{1i}(s),Y_i(s,t_2) \, ; \, 0 \leq s \leq t_1 \leq t_2\} \, .
$$
Since the martingale term is mean-zero noise, the above leads to
\begin{equation}\label{eq:est1}
 \widehat{\Lambda}(dt_1|t_{2}) = \frac{\sum_{i=1}^n \delta_{2i} I(V_{2i}=t_2) dN_{1i}(t_1)}{\sum_{i=1}^n \delta_{2i} I(V_{2i}=t_2)Y_{i}(t_1,t_2)}   \, .  
\end{equation}
An estimator of $S_{1|2}(t_1|t_2)$ is obtained by the product integral of $1-\widehat{\Lambda}_{1|2}(dt_1|t_{2})$,
$$
\widehat{S}_{1|2}(t_1|t_2) = \prod_{s \leq t_1}\left\{ 1-\widehat{\Lambda}_{1|2}(ds|t_{2}) \right\} \, .
$$
Finally,
\begin{equation}\label{eq:final}
\widetilde{G}_1(t_1) = \sum_{j=1}^{J_2}  \{ 1- \widehat{S}_{1|2}(t_1 \wedge t_{2,j}|t_{2,j}) \} \Delta \widehat{F}_2(t_{2,j})
\end{equation}
where $t_{2,j}$, $j=1,\ldots,J_2$, are the ordered  distinct observed death ages,  Clearly, $\widetilde{G}_1$ is an estimator of CIF given that $T_2 \geq c_L$. However, in contrast to $\widehat{G}^{AJ}_1$, owing to prevalent events that may occur at times less than $c_L$, the estimator $\widetilde{G}_1(t)$ is not restricted to $t > c_L$. 

We emphasize that even when the goal is to estimate
$G_1(t|T_1 \geq c_L, T_2 \geq c_L)$, the estimator $\widetilde{G}_1(t)$ based on the subsample  restricted by $T_1 \geq c_L$
only is expected to be more efficient than $\widehat{G}^{AJ}_1$ since some of the prevalent observations (and often the majority of them) 
are with $T_1 \geq c_L$.  

In case of no tied death ages,  $\widehat{\Lambda}_{1|2}(dt_1|T_{2i})=\delta_{1i}\delta_{2i}I(V_{1i}=t_1)$ and therefore
$1-\widehat{S}_{1|2}(t_1|T_{2i})=\delta_{1i}\delta_{2i}I(V_{1i} \leq t_1)$. Hence,
 $\widetilde{G}_1(t)$ reduces to 
\begin{eqnarray}\label{eq:final2}
\widehat{G}_1(t_1)  &=&   \sum_{i=1}^n  \delta_{1i}\delta_{2i}  I(V_{1i} \leq t_1) \Delta \widehat{F}_2(T_{2i}) \nonumber \\
&=& \frac{1}{n} \sum_{i=1}^n \delta_{1i}\delta_{2i}  I(V_{1i} \leq t_1) \frac{\widehat{S}_2(T^-_{2i})}{\bar{Y}_{2.}(T_{2i})}
\end{eqnarray}
where $Y_{2i}(t)=I(R_i \leq t \leq V_{2i})$ and $\bar{Y}_{2.}(t)=n^{-1}\sum_{i=1}^n Y_{i2}(t)$. The estimator formulated in (\ref{eq:final2}) exhibits good performance not only in scenarios without tied death times, but also in cases with a moderate amount of tied death times. Consequently, we adopt (\ref{eq:final2}) as our ultimate proposed estimator, regardless of whether the data exhibit tied death times or not.

Define $\cY_2(v) = E\{Y_{2i}(v)|T_{2i} \geq R_i\}$, $K(v) = {S}_2(v^-|c_L)/ \cY_{2}(v)$, $S_2(t|s)=\Pr(T_2 > t|T_2 >s)$
and $\widehat{K}(v) = \widehat{S}_2(v^-)/\bar{Y}_{2.}(v)$.
We can then write
\begin{align}
\widehat{G}_1(t_1) 
= \frac{1}{n} \sumi \delta_{1i} \delta_{2i}  \widehat{K}(V_{2i}) I(V_{1i} \leq t_1) \, .
\label{gf}
\end{align}
The asymptotic properties of $\widehat{G}_1$ are outlined in the following theorem. The assumptions and the proof of consistency are detailed in Appendix 1, while the proof of asymptotic normality is provided in the Supplementary Material.
\begin{theorem}
If Assumptions A.1 - A.3 hold, as $n \rightarrow \infty$, $$\sup_{t \in [0,\tau]}|\widehat{G}_1(t) - G_1(t|T_2 \geq c_L)| = o_{a.s.}(1)$$
and $\sqrt{n} \{ \widehat{G}_1(t) - G_1(t|T_2 \geq c_L) \}$ converges weakly to a Gaussian process. \\
\end{theorem}

We can also form a combination estimator $\widehat{G}_1^{cmb}(t)$ as $\widehat{G}_1^{cmb}(t) = 0.5\widehat{G}_1^{AJ}(t)+0.5\widehat{G}_1(t)$.

\section{CIF Point-wise Confidence Intervals and Simultaneous Confidence Band}
In the Supplementary Material,  $\widehat{G}_1(t) - G_1(t|T_2 \geq c_L)$ is represented as a mean of independent, identically distributed terms with zero mean plus a negligible remainder. Namely, 
\begin{equation}
\widehat{G}_1(t) - G_1(t|T_2 \geq c_L) = \frac{1}{n} \sum_{i=1}^n \Psi_i(t) + o_p(n^{-1/2}) \, ,
\label{rep}
\end{equation}
where detailed expressions are given in the Supplementary Material for $\Psi_i(t)$ and its estimator $\widehat{\Psi}_i(t)$. 
The estimator $\widehat{\Psi}_i(t)$ is the sum of two terms: a main term, given by
\begin{equation}
\widehat{\Psi}_{1i}(t) = \delta_{1i} \delta_{2i} I(V_{1i} \leq t_1) \widehat{K}(V_{2i}) - \widehat{G}_1(t_1)
\label{mt}
\end{equation}
and an auxiliary term $\widehat{\Psi}_{2i}(t)$ that arises from taking into account the fact that
the estimator (\ref{gf}) involves an estimate $\hat{K}$ of $K$ rather than $K$ itself.
Given the representation (\ref{rep}),
the variance $\Var(\sqrt{n} \, \{\widehat{G}_1(t) - G_1(t|T_2>c_L) \})$ can be estimated using the empirical estimator
\begin{equation*}
s^2(t) = \widehat{\Var}(\sqrt{n} \, \{\widehat{G}_1(t) - G_1(t|T_2>c_L) \}) = 
\frac{1}{n} \sumi \widehat{\Psi}^2_i(t) \, 
\end{equation*}
We can then form a point-wise $100(1-\alpha)\%$ confidence interval for $G_1(t)$ as
$$
\widehat{G}_1(t) \pm \zeta_{1-\alpha/2} \, \frac{s(t)}{\sqrt{n}}, \quad \zeta_{1-\alpha/2}=\Phi^{-1}(1-\alpha/2)	\, .
$$
Alternatively, we can form a confidence interval on a transformed scale, using, for example, the log transformation $g(u)=-\log(1-u)$ or the arcsine-root transformation
$g(u)=\pi/2-\arcsin{\sqrt{1-u}}$. This confidence interval takes the form
$$
g(\widehat{G}_1(t)) \pm \zeta_{1-\alpha/2} \, g^\prime(\widehat{G}_1(t)) \frac{s(t)}{\sqrt{n}} \, ,
$$
and we can then apply the inverse transformation to obtain a confidence interval on the original scale.

To construct a  simultaneous confidence band for $G_1(t|T_2 \geq c_L)$ along the lines of the equal-precision band of \cite{nair1984},
we use the resampling approach of \cite{fleming1994confidence}.
Suppose we want a simultaneous confidence band over the interval $[\tau_1,\tau_2]$.
Let $Z_{bi}$ be independent standard normal random variables, $b=1,\ldots,B$, $i=1,\ldots,n$, and define
\begin{align*}
\Delta^{(b)}(t) & = \frac{1}{n} \sum_{i=1}^n Z_{bi} \widehat{\Psi}_i(t) \, , \\
\Gamma^{(b)}(t) & = s(t)^{-1} \Delta^{(b)}(t) \, .
\end{align*}
Let $M^{(b)} = \max_{t \in [\tau_1,\tau_2]} |\Gamma^{(b)}(t)|$ and define $\nu_{\alpha}$ to be the $1-\alpha$ quantile of $M^{(1)},\ldots,M^{(B)}$.
The confidence band is then given by
$$
\widehat{G}_1(t) \pm \nu_{\alpha} \frac{s(t)}{\sqrt{n}},   \,\,\,\,\,  t \in [\tau_1,\tau_2] \ .
$$
It is also possible to form a confidence band on a transformed scale. Define
\begin{align*}
\check{G}^{(b)}_1(t) & = \widehat{G}_1(t) +\frac{1}{n}\sum_{i=1}^n Z_{bi} \widehat{\Psi}_i(t) - \frac{1}{n} \sum_{i=1}^n  \widehat{\Psi}_i(t) \, , \\ 
\check{\Delta}^{(b)}(t) & = g(\check{G}^{(b)}_1(t)) - g(\hat{G}_1(t)) \, , \\
\check{\Gamma}^{(b)}(t) & =\left (g^\prime(\check{G}^{(b)}_1(t)) s(t) \right)^{-1} \check{\Delta}^{(b)}(t) \, ,\\
\check{M}^{(b)} & = \max_{t \in [\tau_1,\tau_2]} |\check{\Gamma}^{(b)}(t)| \, .
\end{align*}
In addition,  define $\check{\nu}_{\alpha}$ to be the $1-\alpha$ quantile of $\check{M}^{(1)},\ldots,\check{M}^{(B)}$.
The confidence band on the transformed scale is then
$$
\widehat{G}_1(t) \pm \check{\nu}_{\alpha} g^\prime(\widehat{G}_1(t)) \frac{s(t)}{\sqrt{n}},   \,\,\,\,\,  t \in [\tau_1,\tau_2] \ ,
$$
and we can apply the inverse transformation to obtain a simultaneous confidence band on the original scale.

A similar confidence band procedure can be developed for the Aalen-Johansen estimator; see the Supplementary Materials for details.

\section{Simulation Study}
We conducted an extensive simulation study to examine the finite-sample properties of the proposed estimator, $\widehat{G}_1$, in comparison with the Aalen-Johansen estimator, $\widehat{G}_1^{AJ}$, adjusted for left truncation. Table \ref{tbl:sim-desc} summaries the configurations examined. Each configuration was studied with 1,000 repetitions and 250 bootstrap samples, with sample sizes of 2,500, 5,000 and 7,500. The arcsine-root transformation was used for the point-wise confidence intervals and simultaneous confidence bands.

In preliminary investigations, we found that the variance estimator based only on the main term (\ref{mt}) produced estimates similar to those obtained using both the main
term and the auxiliary terms. Accordingly, to expedite computation, the final simulation results exclude the auxiliary terms from the variance estimator. The R code implementing
the methods give the user the option to include or exclude the auxiliary terms.

\begin{figure}
\centering
\includegraphics[width=1.05\textwidth]{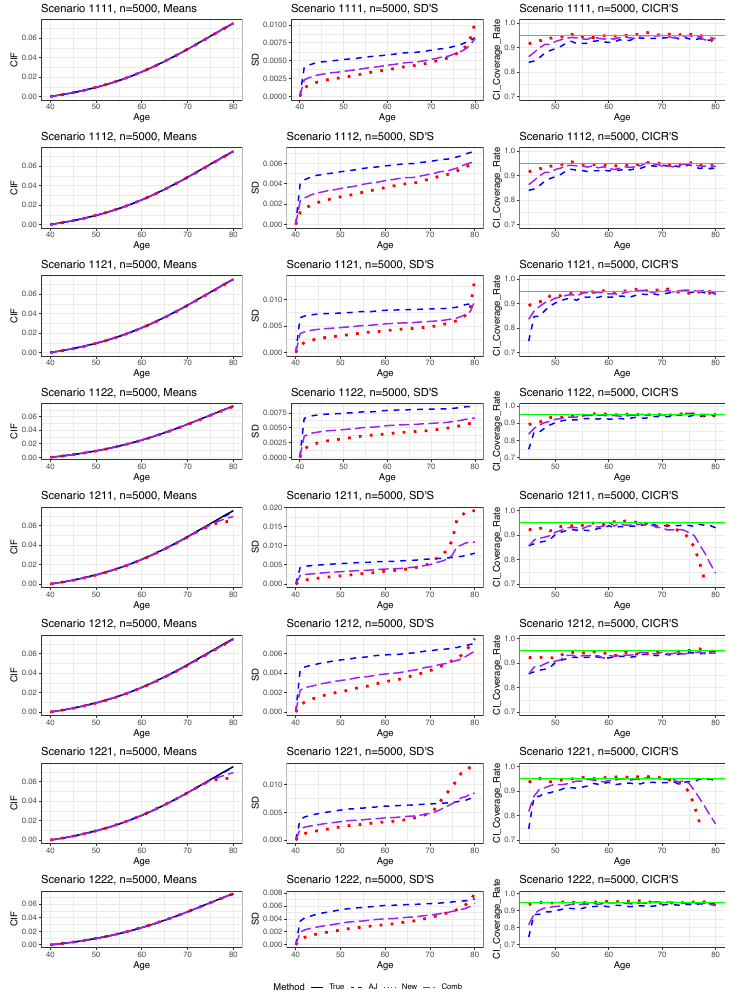}
\caption{Simulation results of eight configurations: Mean over estimates, standard deviations (SD), and empirical coverage rates of 95\% point-wise confidence intervals, for each of the three methods, AJ, the new estimator, and the combination estimator (Comb). \label{fig:sim16A} }
\end{figure}

\begin{figure}
\centering
\includegraphics[width=1.05\textwidth]{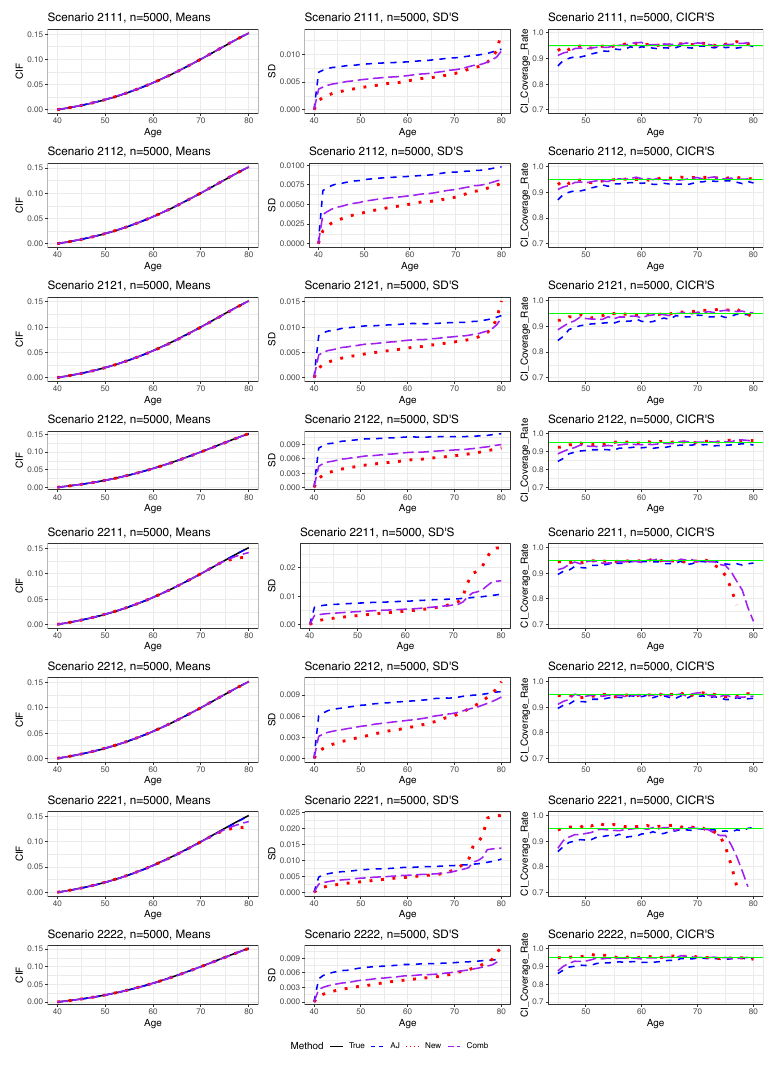}
\caption{Simulation results of eight configurations: Mean over estimates, standard deviations (SD), and empirical coverage rates of 95\% point-wise confidence intervals, for each of the three methods, AJ, the new estimator, and the combination estimator (Comb). \label{fig:sim16B} }
\end{figure}

Figures \ref{fig:sim16A} and \ref{fig:sim16B} summarize  the results of 16 configurations in which the disease hazard is zero up until the minimum recruitment age of 40 years. The plots show the results of three estimators, the Aalen-Johansen estimator, the new proposed estimator, and the combination estimator, with $n=5,000$ observations, for ages between 40 and 80.
The estimated standard deviation of the Aalen-Johansen estimator was obtained using the function \texttt{etmCIF} in the R package \texttt{etm}; estimates obtained using the approach used for the new proposed estimator
yielded essentially identical results.
The graphs in the figures present the mean, 
empirical standard deviation, and point-wise confidence interval coverage of the estimators over the 1,000 simulation replications, as a function of $t$.
The Supplementary Material provides additional information, including median, interquartile range,
comparison of mean estimated standard deviation and the empirical standard deviation, confidence interval widths, and confidence band widths.
The Supplemental Materials also include results for $n=2,500$ and $n=7,500$.

In general, all the three estimators are similarly well-behaved in terms of bias. A slight downward bias was seen with
the new estimator at the upper end of the age range (ages 75 to 80) for Scenarios 1211, 1221, 2211, and 2221. These were scenarios which involved both a short duration of follow-up
and a long length of time between disease diagnosis and death.
The new estimator tends to yield a smaller standard deviation than the Aalen-Johansen estimator over the age range from 40 to 75. An upturn in the standard deviation was seen occasionally with the new estimator in the age range from 75 to 80. The standard deviation with the combination estimator was similar, although
somewhat higher, to that with the new estimator over most of the age range. The upturn in the standard deviation of the new estimator at the upper
end of the age range was dampened by the combination estimator.  

The empirical coverage rates of the three methods are usually reasonably close to the nominal level of $95\%$, with some advantage seen with the new method. For Scenarios 1211, 1221, 2211, and 2221, the confidence interval coverage with the new estimator
was substantially lower than nominal in the upper end of the age range, in line with the downward bias previously noted.

\begin{figure}
\centering
\includegraphics[width=0.8\textwidth]{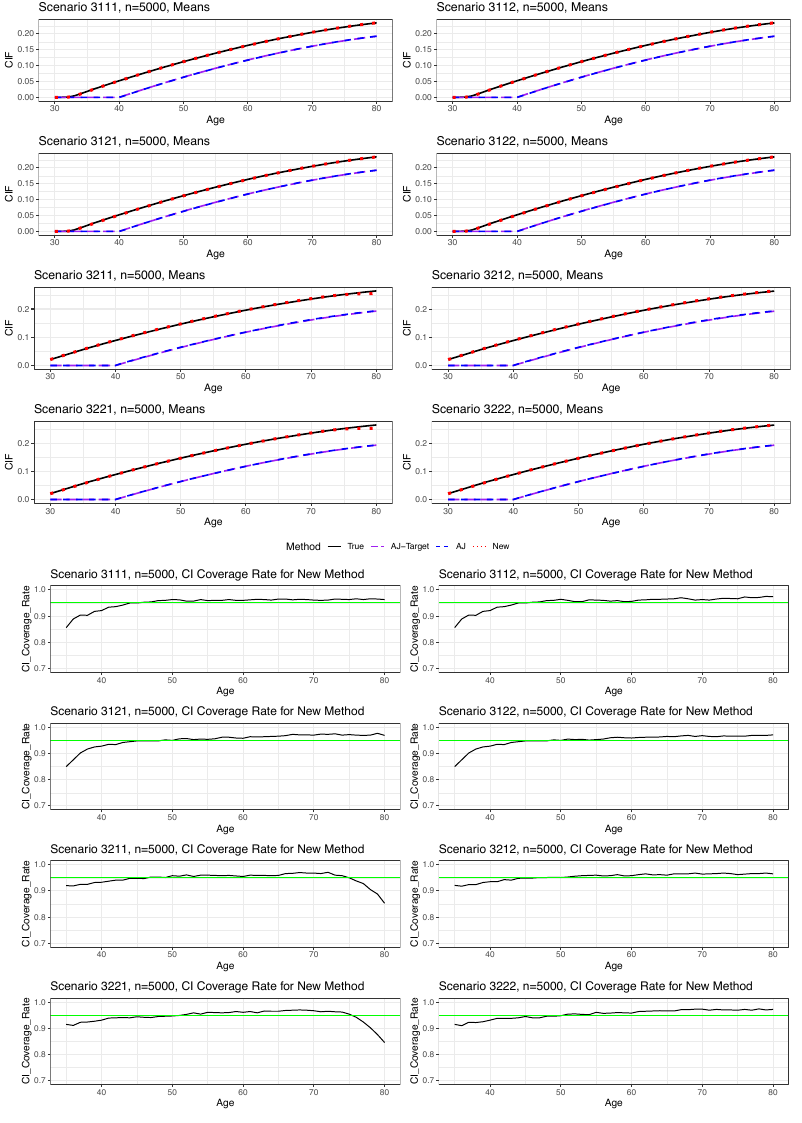}
\caption{Simulation results of 8 configurations: Mean over estimates, standard deviations (SD), and empirical coverage rates of 95\% point-wise confidence intervals, for AJ and the new estimator. \label{fig:sim8} }
\end{figure}

Figure \ref{fig:sim8} provides a similar summary of simulation results, but for eight settings where the disease hazard before the minimum recruitment age of 40 is positive. Clearly, the Aalen-Johansen estimator and the new estimator are targeting different survival curves. Hence, empirical coverage rates are presented only for the proposed approach. Again, the proposed approach performs well in terms of bias and point-wise coverage rates.

Table \ref{tbl:sim-24bands} presents the empirical coverage rates of the 95\% simultaneous confidence bands.
For Scenarios 1xxx and 2xxx, the confidence band was computed over the age range 50-80, except for Scenarios 1211, 1221, 2211, and 2221, for which the
band was computed over the age range 50-75. For Scenarios 3xxx, the age range was 35-80.
Although all estimators show excellent performance in terms of bias and point-wise coverage rates already with $n=2,500$ (see the Supplemental
Materials), achieving adequate coverage rates for the confidence bands necessitates a larger sample size. In certain settings, the coverage rates are reasonably close to 0.95, yet in others, a larger sample size is essential for satisfactory coverage. Notably, the new estimator often surpasses the Aalen-Johansen estimator in terms of empirical coverage rates.

\section{UK Biobank Data Analysis}
The  Aalen-Johansen estimator and the proposed estimator were applied for two types of cancers, acute myeloid
leukemia (AML) and brain cancer, separately for males and females. Since the UKB recruited volunteers aged 40 to 69, the Aalen-Johansen CIF estimator begins at age 40. The Aalen-Johansen estimator estimates the CIF given being alive and free of the disease by age 40. The new approach provides a CIF estimator given being alive by age 40, facilitating straightforward interpretation for non-lethal diseases before the age of 40, like AML and most types of brain cancer.

Disease failure times were determined using the first record of ICD9 and ICD10 codes. Censoring occurred at loss to follow-up or being healthy at recent dataset update. To ensure anonymity, UKB only reports birth month and year, excluding exact birth dates, except for cancer diagnoses or deaths where exact birth dates were inferred from available data. If this was not feasible, birth dates were assigned as the first of the reported birth month.

Table \ref{tbl:UKB-samplesize} summarizes the sample sizes, number of cases, number of prevalent cases, and number of incident cases. Clearly, the majority of the prevalent events occurred at an onset age beyond 40. Therefore, even an analysis that conditions on being alive and free of the disease by age 40, 
can benefit from the proposed approach.
The  prevalent event proportion among all observations diagnosed with the disease ranges from 13\% to 26\%. Additionally, the minimum ages at onset range from 25 to 37 years among prevalent events and from 42 to 46 years among incident events, with differences between 8 to 17 years. 

\begin{figure}
\centering
\includegraphics[width=0.65\textwidth]{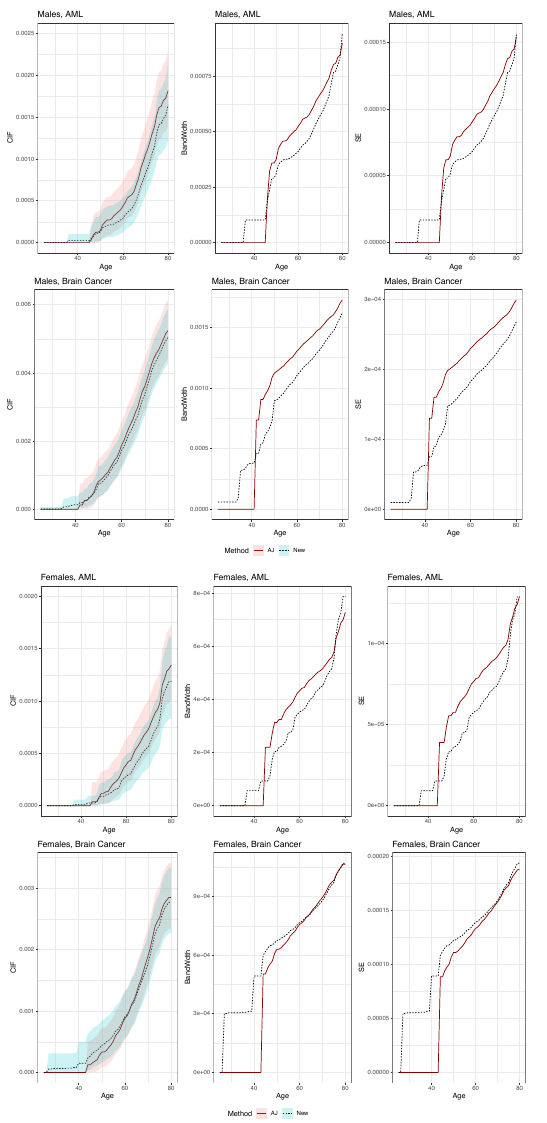}
\caption{UK Biobank Data: analysis of acute myeloid leukemia (AML) and Brain cancer, by sex. Within each sex group, the CIF estimates based on AJ and the new method are provided, along with confidence bands. The bandwidths of the confidence bands and point-wise standard error as a function of age are presented alongside.  \label{fig:AMLBrain} }
\end{figure}

Figure \ref{fig:AMLBrain} displays the CIF estimates and confidence bands, including the widths of the confidence bands and point-wise standard errors, all as functions of age. Clearly, the CIF estimates from both methods are very similar, yet in most instances, the proposed approach yields notably smaller point-wise standard errors and narrower confidence bands. For instance, in the case of AML in males, the confidence band produced by the proposed method is up to 27\% narrower compared to the one from the Aalen-Johansen estimator.
 
\section{Concluding Remarks}

In this paper, we introduce a new CIF estimator that effectively utilizes prevalent data. Its utility is demonstrated through an extensive simulation study and analysis of UK Biobank data, which features a high rate of prevalent events, as is commonly expected in population-based biobanks. While we have demonstrated its advantages over the Aalen-Johansen CIF estimator, it is important to highlight a limitation: for diseases with a relatively young onset age and very low excess mortality rate, such as breast cancer, our proposed estimator is currently inapplicable.  This limitation stems from the current lack of sufficient data to estimate  $S(t_1|t_2)$. In particular, in the UKB breast cancer data, there are 8,729 prevalent case and 8,731 incident cases, among the 17,458  breast cancer cases, but currently, only 2,233 died after breast cancer diagnosis. However, as longer follow-up periods become available in the future, our proposed estimator will facilitate the inclusion of the 8,729 prevalent events and enable the estimation of the breast cancer survival curve before the age of 40. In practical terms, if $\widehat{G}_1$ is substantially lower than  $\widehat{G}_1^{AJ}$, it may indicate that there is insufficient data for properly estimating  $S(t_1|t_2)$.

R code for the estimation procedure and simulations can be found at  

\url{https://github.com/david-zucker/illness-death}.

\section{Acknowledgements}
We thank Bella Vakulenko-Lagun for helpful discussions.
The work of MG is supported in part by  the Israel Science Foundation (ISF) grant
number 767/21 and by a grant from the Tel-Aviv University Center for AI and Data Science
(TAD).
This research has
been conducted using the UK Biobank Resource, project 56885.

\clearpage
\bibliographystyle{abbrvnat}
\bibliography{arxiv}

\begin{thebibliography}{}

\bibitem[\protect\citeauthoryear{Aalen and Johansen}{Aalen and
  Johansen}{1978}]{aalen1978empirical}
Aalen, O.~O. and S.~Johansen (1978).
\newblock An empirical transition matrix for non-homogeneous markov chains
  based on censored observations.
\newblock {\em Scandinavian Journal of Statistics\/}, 141--150.

\bibitem[\protect\citeauthoryear{Allignol, Schumacher, and Beyersmann}{Allignol
  et~al.}{2010}]{allignol2010note}
Allignol, A., M.~Schumacher, and J.~Beyersmann (2010).
\newblock A note on variance estimation of the aalen--johansen estimator of the
  cumulative incidence function in competing risks, with a view towards
  left-truncated data.
\newblock {\em Biometrical Journal\/}~{\em 52\/}(1), 126--137.

\bibitem[\protect\citeauthoryear{Chang and Tzeng}{Chang and
  Tzeng}{2006}]{chang2006nonparametric}
Chang, S.-H. and S.-J. Tzeng (2006).
\newblock Nonparametric estimation of sojourn time distributions for truncated
  serial event data—a weight-adjusted approach.
\newblock {\em Lifetime data analysis\/}~{\em 12\/}(1), 53--67.

\bibitem[\protect\citeauthoryear{Gill and Johansen}{Gill and
  Johansen}{1990}]{gill1990}
Gill, R.~D. and S.~Johansen (1990).
\newblock A survey of product-integration with a view toward application in
  survival analysis.
\newblock {\em Annals of Statistics\/}~{\em 18\/}(4), 1501--1555.

\bibitem[\protect\citeauthoryear{Keiding}{Keiding}{1991}]{keiding1991age}
Keiding, N. (1991).
\newblock Age-specific incidence and prevalence: a statistical perspective.
\newblock {\em Journal of the Royal Statistical Society: Series A (Statistics
  in Society)\/}~{\em 154\/}(3), 371--396.

\bibitem[\protect\citeauthoryear{Kosorok}{Kosorok}{2008}]{kosorok2008}
Kosorok, M.~R. (2008).
\newblock {\em Introduction to empirical processes and semiparametric
  inference}.
\newblock Springer.

\bibitem[\protect\citeauthoryear{Lin, Fleming, and Wei}{Lin
  et~al.}{1994}]{fleming1994confidence}
Lin, D., T.~Fleming, and L.~Wei (1994).
\newblock Confidence bands for survival curves under the proportional: Hazards
  model.
\newblock {\em Biometrika\/}~{\em 81\/}(1), 73--81.

\bibitem[\protect\citeauthoryear{Nair}{Nair}{1984}]{nair1984}
Nair, V.~N. (1984).
\newblock Confidence bands for survival functions with censored data: a
  comparative study.
\newblock {\em Technometrics\/}~{\em 26\/}(3), 265--275.

\bibitem[\protect\citeauthoryear{Saarela, Kulathinal, and Karvanen}{Saarela
  et~al.}{2009}]{saarela2009joint}
Saarela, O., S.~Kulathinal, and J.~Karvanen (2009).
\newblock Joint analysis of prevalence and incidence data using conditional
  likelihood.
\newblock {\em Biostatistics\/}~{\em 10\/}(3), 575--587.

\bibitem[\protect\citeauthoryear{Tsai}{Tsai}{1990}]{tsai1990testing}
Tsai, W.-Y. (1990).
\newblock Testing the assumption of independence of truncation time and failure
  time.
\newblock {\em Biometrika\/}~{\em 77\/}(1), 169--177.

\bibitem[\protect\citeauthoryear{Vakulenko-Lagun and Mandel}{Vakulenko-Lagun
  and Mandel}{2016}]{vakulenko2016comparing}
Vakulenko-Lagun, B. and M.~Mandel (2016).
\newblock Comparing estimation approaches for the illness--death model under
  left truncation and right censoring.
\newblock {\em Statistics in Medicine\/}~{\em 35\/}(9), 1533--1548.

\bibitem[\protect\citeauthoryear{Vakulenko-Lagun, Mandel, and
  Goldberg}{Vakulenko-Lagun et~al.}{2017}]{vakulenko2017nonparametric}
Vakulenko-Lagun, B., M.~Mandel, and Y.~Goldberg (2017).
\newblock Nonparametric estimation in the illness-death model using prevalent
  data.
\newblock {\em Lifetime Data Analysis\/}~{\em 23\/}(1), 25--56.

\bibitem[\protect\citeauthoryear{van~der Vaart}{van~der Vaart}{1998}]{vdv1998}
van~der Vaart, A. (1998).
\newblock {\em Asymptotic statistics}.
\newblock Cambridge University Press.

\bibitem[\protect\citeauthoryear{Xu, Kalbfleisch, and Tai}{Xu
  et~al.}{2010}]{xu2010statistical}
Xu, J., J.~D. Kalbfleisch, and B.~Tai (2010).
\newblock Statistical analysis of illness--death processes and semicompeting
  risks data.
\newblock {\em Biometrics\/}~{\em 66\/}(3), 716--725.

\end{thebibliography}

\vspace*{-10pt}

\begin{table}[]
\scriptsize
\centering
\caption{Summary of simulation settings: Age at death is sampled based on the UK Office of National Statistics}\label{tbl:sim-desc}
\begin{tabular}{lcc}
  \hline
Random Variable & Distributions & Setting Notation \\ 
     \hline
     Age at disease diagnosis, $T_1$ & Truncated Weibull with shape=4, scale=115, truncation at 40 & $(1,\cdot,\cdot,\cdot)$ \\
      & Truncated Weibull with shape=4, scale=130, truncation at 40 & $(2,\cdot,\cdot,\cdot)$\\
      & Weibull with shape=3.5, scale=200 & $(3,\cdot,\cdot,\cdot)$\\
      \hline
     Age at death, $T_2$   
     & expected survival after disease diagnosis equals 2.5 years for Setting 1 or 2 of $T_1$  & $(1 \mbox{ or } 2,1,\cdot,\cdot)$ \\ 
     & expected survival after disease diagnosis equals 7.5 years  for Setting  1 or 2 of $T_1$ & $(1 \mbox{ or } 2,2,\cdot,\cdot)$\\
      & expected survival after disease diagnosis equals 5 years for Setting 3 of $T_1$   & $(3,1,\cdot,\cdot)$ \\
      & expected survival after disease diagnosis equals  10 years for Setting  3 of $T_1$ & $(3,2,\cdot,\cdot)$ \\
\hline
    Recruitment age, $R$  & Uniform distribution over $[40,69]$ & $(\cdot,\cdot,1,\cdot)$ \\
      & The recruitment distribution of the UKB & $(\cdot,\cdot,2,\cdot)$ \\
 \hline     
      Censoring age, $C$ & Recruitment age + uniform over $[11,15]$ & $(\cdot,\cdot,\cdot,1)$\\
                        & Recruitment age + uniform over $[11,25]$ & $(\cdot,\cdot,\cdot,2)$\\
      \hline
\end{tabular}
\end{table}

\begin{table}[]
\scriptsize
\centering
\caption{Summary of simulation results: empirical coverage rates of the 95\% confidence-band }\label{tbl:sim-24bands}
\begin{tabular}{l|ccc|ccc|ccc}
  \hline
   & \multicolumn{3}{c}{$n=2500$}  & \multicolumn{3}{c}{$n=5000$} & \multicolumn{3}{c}{$n=7500$}  \\
  \hline
Setting  & AJ & New & Combo & AJ & New & Combo & AJ & New & Combo \\ 
\hline
1111 & 0.901 & 0.912 & 0.900 & 0.919 & 0.923 & 0.923 & 0.937 & 0.916 & 0.931 \\
1112 & 0.906 & 0.939 & 0.911 & 0.922 & 0.939 & 0.925 & 0.937 & 0.933 & 0.940 \\
1121 & 0.883 & 0.900 & 0.900 & 0.914 & 0.919 & 0.928 & 0.909 & 0.925 & 0.913 \\
1122 & 0.886 & 0.911 & 0.899 & 0.915 & 0.935 & 0.932 & 0.902 & 0.919 & 0.914 \\
1211 & 0.915 & 0.901 & 0.921 & 0.914 & 0.898 & 0.905 & 0.897 & 0.905 & 0.920 \\
1212 & 0.920 & 0.923 & 0.926 & 0.922 & 0.924 & 0.919 & 0.895 & 0.941 & 0.927 \\
1221 & 0.895 & 0.897 & 0.880 & 0.916 & 0.904 & 0.922 & 0.908 & 0.908 & 0.918 \\
1222 & 0.894 & 0.921 & 0.899 & 0.914 & 0.929 & 0.927 & 0.912 & 0.939 & 0.925 \\
2111 & 0.919 & 0.912 & 0.926 & 0.939 & 0.945 & 0.956 & 0.934 & 0.947 & 0.934 \\
2112 & 0.923 & 0.940 & 0.934 & 0.932 & 0.951 & 0.951 & 0.929 & 0.951 & 0.946 \\
2121 & 0.919 & 0.924 & 0.926 & 0.918 & 0.938 & 0.935 & 0.932 & 0.916 & 0.937 \\
2122 & 0.924 & 0.940 & 0.930 & 0.916 & 0.946 & 0.936 & 0.934 & 0.948 & 0.942 \\
2211 & 0.931 & 0.920 & 0.924 & 0.937 & 0.899 & 0.932 & 0.930 & 0.910 & 0.922  \\
2212 & 0.929 & 0.941 & 0.931 & 0.932 & 0.936 & 0.937 & 0.934 & 0.944 & 0.940 \\
2221 & 0.909 & 0.898 & 0.906 & 0.928 & 0.918 & 0.911 & 0.916 & 0.901 & 0.921 \\
2222 & 0.907 & 0.934 & 0.919 & 0.931 & 0.946 & 0.931 & 0.927 & 0.948 & 0.928 \\
3111 & - & 0.864 & - & - & 0.900 & - & - & 0.904 & - \\
3112 & - & 0.861 & - & - & 0.905 & - & - & 0.902 & -\\
3121 & - & 0.770 & - & - & 0.875 & - & - & 0.894 & -\\
3122 & - & 0.769 & - & - & 0.873 & - & - & 0.892 & -\\
3211 & - & 0.923 & - & - & 0.937 & - & - & 0.936 & -\\
3212 & - & 0.921 & - & - & 9.936 & - & - & 0.936 & -\\
3221 & - & 0.910 & - & - & 0.926 & - & - & 0.930 & -\\
3222 & - & 0.911 & - & - & 0.926 & - & - & 0.927 & -\\
      \hline
\end{tabular}
\end{table}

\setlength{\tabcolsep}{2pt}
\begin{table}[]
\scriptsize
\centering
\caption{UK Biobank Data: Number of participants alive or died with and without the disease, number of prevalet event before and after age 40, number of incident events, and the minimum observed age at onset among the prevalent and incident participants.  }\label{tbl:UKB-samplesize}
\begin{tabular}{ll|cccc|ccc|cc}
 &         & Alive without & Died without & Alive with & Died with &    Number of  &    Number of       &    Number of       & \multicolumn{2}{c}{Minimal onset age} \\
Sex & Disease & disease       & disease      &disease       & disease & prev $<40$  &  prev $\geq 40$ & incidents & Prevalents & Incidents \\
  \hline
Male & AML & 208,259 & 20,629 & 54 & 147 & 7 & 25 & 169 & 36 & 46 \\
      & Brain & 208,236 & 20,322 & 73 & 458 & 27 & 41 & 462 & 25 & 42 \\
Female & AML &259,038 & 14,111 & 69 & 113 & 14 & 33 & 135 &   37 & 45 \\    
       & Brain & 259,024 & 13,937 & 77 & 293 & 21 & 37 & 312 & 27 & 44 \\
\hline
\end{tabular}
\end{table}

\newpage
\appendix

\section*{Appendix 1: Consistency Proof}
For ease of presentation, the proof is presented for the setting where $R_i, T_{1i}, T_{2i}$, and $C_i$ are all absolutely continuous. We will use the letters $F$, $S$ and $f$ with subscripts to denote cumulative distribution, survival and density functions.

For the asymptotic theory, we make the following standard assumptions. 
\begin{enumerate}
    \item[A.1]  $C_i$ is independent of $(T_{1i}, T_{2i})$.
    \item[A.2] $R_i$ is independent of $T_{1i}$ and quasi-independent of $T_{2i}$ and $C_i$. Namely, for $r$, $t_2$, and $c$ with $c_L~\leq~r\leq~t_2$ and $ c_L~\leq r~\leq c$,  we have 
        $$\Pr (R_i \leq r, T_{2i}>t_2, C_i > c|T_{2i} \geq R_i, C_i \geq R_i) = a^{-1} F_R(r)S_{2}(t_2|c_L)S_C(c)$$
    where $S_2(t_2|c_L)=\Pr(T_2>t_2|T_2>c_L)$ and
    $a=\int \int \int_{r \leq t_2 \wedge c} dF_R(r) dS_{2}(t_2|c_L) dS_C(c)$.
    \item[A.3] $\Pr (Y_{2i}(\tau)=1|T_{2i} \geq R_i)>0$ where $Y_{2i}(t)=I(R_i \leq t)I(V_{2i} \geq t)$.
\end{enumerate}

Our starting point for the consistency proof is the representation (\ref{gf}) of the estimator:
$$
\widehat{G}_1(t_1) 
= \frac{1}{n} \sumi \delta_{1i} \delta_{2i}  \widehat{K}(V_{2i}) I(V_{1i} \leq t_1) 
$$
Using Assumptions A.1--A.2, we have
\begin{align}
K(v) & = \frac{S_2(v-|c_L)}{P(R_i \leq v, V_{2i} \geq v|T_{2i} \geq R_i)}  \nonumber \\
& = \frac{S_2(v-|c_L)}{P(R_i \leq v, T_{2i} \geq v, C_i \geq v | T_{2i} \geq R_i)} \nonumber \\
& = \frac{a}{F_R(v)S_C(v^-)}	\label{a} 
\end{align}
We thus see that the proposed estimator has the form of an IPW estimator, but it is different from the IPW estimators of \cite{chang2006nonparametric} and \cite{vakulenko2017nonparametric} (applicable for $c_L>0$ and much simpler). Write 
\begin{align*}
H_i(t_1) & = \delta_{2i} K(T_{2i}) I(T_{1i} \leq T_{2i})I(T_{1i} \leq t_1) = \delta_{1i} \delta_{2i} K(V_{2i}) I(V_{1i} \leq t_1) \\
\widehat{H}_i(t_1) & = \delta_{2i} \widehat{K}(T_{2i}) I(T_{1i} \leq T_{2i})I(T_{1i} \leq t_1) = \delta_{1i} \delta_{2i} \widehat{K}(V_{2i}) I(V_{1i} \leq t_1)  \, .
\end{align*}
Our estimator can be written as
\begin{equation}
\widehat{G}_1(t_1) = \frac{1}{n} \sum_{i=1}^{n} \widehat{H}_i(t_1) \, .
\end{equation}
Defining $\| c \|_\infty = \sup_{t \in [0,\tau]} |c(t)|$, it follows from the 
Glivenko-Cantelli theorem that $\|\bar{Y}_{2n} - \cY_2\|_{\infty} = o_{a.s.}(1)$ and from known theory for the Kaplan-Meier estimator that  
$\| \widehat{S}_2 - S_2 \|_{\infty} = o_{a.s.}(1)$. Consequently,
\begin{equation}
\widehat{G}_1(t_1) = \frac{1}{n} \sumi {H}_i(t_1) + o_{a.s.}(1)
\end{equation}
In the Supplementary Material,  an expansion of the remainder term is presented. Meantime, from the above equation, 
we see that $\widehat{G}_1(t_1)$ converges almost surely to $E[H_i(t_1)|T_{2i} \geq R_i]$.
We now show that $E[H_i(t_1)|T_{2i} \geq R_i] = G_1(t_1)$.  We have
\begin{align*}
& E[H_i(t_1)|T_{2i} \geq R_i] \\
& \hspace*{3mm} = a^{-1} \int \int f_{(T_1,T_2|T_2>c_L)}(t_1^\circ,t_2|T_2>c_L) K(t_2)I(t_1^\circ \leq t_2)I(t_1 \leq t_1^\circ) \\
& \hspace*{12mm} 
\left\{ \int f_{R}(r) I(r \leq t_2) dr \int f_C(c) I(C \geq t_2) dc \right\} dt_1^\circ dt_2 \\
& \hspace*{3mm} = a^{-1} \int \int f_{(T_1,T_2|T_2>c_L)}(t_1^\circ,t_2|T_2>c_L)K(t_2)I(t_1^\circ \leq t_2)I(t_1 \leq t_1^\circ) 
\Pr(R_i \leq t_2) \Pr( C_i \geq t_2) dt_1 ^\circ dt_2 \\
& \hspace*{3mm} = a^{-1} \int \int f_{(T_1,T_2|T_2>c_L)}(t_1^\circ,t_2|T_2>c_L) \left\{ \frac{a}{F_R(t_2)S_C(t_2^-)} \right\}
I(t_1^\circ \leq t_2)I(t_1^\circ \leq t_1) \\
& \hspace*{40mm}\Pr(R_i \leq t_2) \Pr( C_i \geq t_2)  dt_1^\circ dt_2 \\
& \hspace*{3mm} (\mbox{using \ref{a}}) \\
& \hspace*{3mm} = \int \int f_{(T_1,T_2|T_2>c_L)}(t_1^\circ,t_2|T_2>c_L) I(t_1^\circ \leq t_2)I(t_1^\circ \leq t_1) dt_1^\circ dt_2 \\
& \hspace*{3mm} = G_1(t_1|T_2>c_L)
\end{align*}
We have thus shown that $\widehat{G}_1(t_1)$ is an almost-sure consistent estimator of $G_1(t_1|T_2>c_L)$.


\section*{Supplementary Material}

\section*{Proof of Theorem 1 - Asymptotic Normality} 

\vspace*{3mm}

Recall the expression for $\widehat{G}_1(t_1)$:
$$
\widehat{G}_1(t_1)=
\frac{1}{n} \sumi \delta_{1i} \delta_{2i} \left\{ \frac{\widehat{S}_2(V_{2i}-)}
{\bar{Y}_{2n}(V_{2i})} \right\} I(V_{1i} \leq t_1) = \frac{1}{n} \sumi \widehat{H}_i(t_1) 
$$
If we knew $S_2$ and $\cY_2$, we would write
\begin{equation}
\widehat{G}_1(t_1)=
\frac{1}{n} \sumi \delta_{1i} \delta_{2i} \left\{ \frac{{S}_2(V_{2i}-)}
{\cY_2(V_{2i})} \right\} I(V_{1i} \leq t_1) = \frac{1}{n} \sumi {H}_i(t_1) 
\label{gs}
\end{equation}
In this case, $\widehat{G}_1(t_1)$ would be a simple average of i.i.d.\ random variables, asymptotically normal by the classical central limit theorem,
with variance that could be estimated using the empirical estimator
\begin{equation}
\widehat{\Var}(\widehat{G}_1(t_1)) = \frac{1}{n} \left(\frac{1}{n} \sumi (H_i(t_1) - \bar{H}(t_1))^2 \right)
=  \frac{1}{n} \left(\frac{1}{n} \sumi (H_i(t_1) - \widehat{G}_1(t_1))^2 \right)
\end{equation}
With $S_2$ and $\cY_2$ unknown, we plug in the estimates $\widehat{S}_2$ and $\bar{Y}_{2n}$. We might then use the variance estimator
\begin{equation}
\widehat{\Var}(\widehat{G}_1(t_1)) = \frac{1}{n} \left(\frac{1}{n} \sumi (\widehat{H}_i(t_1) - \widehat{G}_1(t_1))^2 \right)
\label{var}
\end{equation}
This estimator, however, would not be entirely correct since it doesn't account for the variability due to
estimation of $S_2$ and $\cY_2$.

A more complete characterization of the asymptotic behaviour of $\widehat{G}_1(t_1)$ is presented below. The development involves 
representing $\widehat{G}_1(t_1^\circ)$ as the mean of i.i.d.\ 
quantities up to an error of $o_P(n^{-1/2})$. We will develop this representation using empirical process theory and 
the functional delta method (Chapters 19 and 20 of \cite{vdv1998}).

We remark that in our numerical work we found that accounting for the estimation error in $S_2$ and $\cY_2$
leads to a generally modest (and often negligble) difference relative to estimating the variance using (\ref{var}).

Define $\bar{N}_{2n}(v) = n^{-1} \sumi N_{2i}(v)$ and $\cN_2(t) = E[N_{2i}(v) | T_{2i} \geq R_i]$. Also define
\begin{align*}
Q_n(v;t_1) & = \frac{1}{n} \sumi \delta_{1i} \delta_{2i} I(V_{1i} \leq t_1) I(V_{2i} \leq v) \\
Q(v;t_1) & = E[\delta_{1i} \delta_{2i} I(V_{1i} \leq t_1) I(V_{2i} \leq v)]	
\end{align*}
We can then write
\begin{equation}
\widehat{G}_1(t_1) = \int_0^\tau \widehat{K}(v) \, dQ_n(v;t_1)
= \int_0^\tau \frac{\widehat{S}_{2\didl} (v-)}{\bar{Y}_{2n}^{\didl} (v)} \, dQ_n(v;t_1)
\end{equation}
and
\begin{equation}
\widehat{S}_2(v) = \msp_{[0,v]} \left( 1 - \frac{d\bar{N}_{2n}(\tv)}{\bar{Y}_{2n}(\tv)} \right)	
\end{equation}
where $\msp$ denotes the product integral. We will regard $\bar{N}_{2n}(v)$ as a stochastic process taking values in the space
$D[0,\tau]$ of functions on $[0,\tau]$ that are right-continuous with left limits and $\bar{Y}_{2n}(v)$ as a stochastic process
taking values in the space $D_-[0,\tau]$ of functions on $[0,\tau]$ that are left-continuous with right limits.
We will use the abbreviations $D$ and $D_-$ for these spaces. We endow $D$ and $D_-$ with the uniform norm 
$\|\cdot \|_{\infty}$. In addition, we regard $Q_n(v;t_1)$ as a stochastic process taking values in 
the space $\cH$ of functions $b(v;t_1)$ whose total variation with respect to $v$ is bounded by 1 for every $t_1$,
and we endow this space with the norm $\|b\|_{\cH}$ given by the supremum over $t_1$ of the total variation of $b$
of $b(v;t_1)$ with respect to $v$.
We will write $\cG = D \times D_- \times \cH$. 

We now introduce the following definitions:
\begin{align*}
& \cM_1: \cG \rightarrow \cG
\mbox{ defined as }
\cM_1(\cA_1,\cA_2,\cA_3) = (\cA_1,\cA_2^{-1},\cA_3) \\
& \cM_2: \cG \rightarrow \cG
\mbox{ defined as }
\cM_2(\cB_1,\cB_2,\cB_3) = 
\left( \int_{[0,\cdot]} B_2 \, dB_1, \cB_2, \cB_3 \right) \\
& \cM_3: \cG \rightarrow \cG
\mbox{ defined as }
\cM_3(\cC_1,\cC_2, \cC_3) = 
( \msp_{[0,\cdot]} (1 - d \cC_1), \cC_2, \cC_3) \\
& \cM_4: \cG \rightarrow D
\mbox{ defined as }
\cM_4(\cD_1,\cD_2,\cD_3) = \int_0^\tau \cD_1(v-) \cD_2(v) \, d\cD_3(v;\cdot)
\end{align*}
Defining $\cM$ to be the composition of all four of the above maps, we have $\widehat{G}_1(\cdot) = 
\cM(\bar{N}_2,\bar{Y}_2,Q_n)$.

From the development in \cite{gill1990}, we see that the first three maps
are Hadamard differentiable with the following derivatives:
\begin{align*}
& \cM_1^\pr(\alpha_1,\alpha_2,\alpha_3|\cA_1,\cA_2,\cA_3)
= (\alpha_1,-\cA_2^{-2}\alpha_2,\alpha_3) \\
& \cM_2^\pr(\beta_1,\beta_2,\beta_3|\cB_1,\cB_2,\cB_3)
= \left( \int_{[0,\cdot]} (\cB_2 \, d\beta_1 + \beta_2 \, d\cB_1), \beta_2, \beta_3 \right) \\
& \cM_3^\pr(\gamma_1,\gamma_2,\gamma_3|\cC_1,\cC_2,\cC_3)
= \left( - \cC_1^* \int_{[0,\cdot]} \frac{\cC_{1-}^*}{\cC_1^*} \, d\gamma_1, \gamma_2, \gamma_3 \right),
\quad \cC_1^* = \msp_{[0,\cdot]} (1-d\cC_1)
\end{align*}
Further, by arguments similar to those in the proofs of Lemma 20.10 of \cite{vdv1998}
and 
Lemma 12.3 of \cite{kosorok2008}, we find that $\cM_4$ is Hadamard differentiable with derivative
\begin{align}
\cM_4^\pr(\epsilon_1,\epsilon_2,\epsilon_3|\cD_1,\cD_2,\cD_3)
& = \int_0^\tau \epsilon_1(v-) \cD_2(v) \, d\cD_3(v;\cdot)
+ \int_0^\tau \epsilon_2(v) \cD_1(v-) \, d\cD_3(v;\cdot) \nonumber \\
& \hspace{10mm} + \int_0^\tau \cD_1(v-) \cD_2(v) \, d\epsilon_3(v;\cdot)
\end{align}
Consequently, applying the chain rule, $\cM$ is Hadamard differentiable at $(\cN_2,\cY_2,Q)$ with derivative
\begin{equation}
\cM^\pr(\alpha_1,\alpha_2,\alpha_3|\cN_2,\cY_2,Q) = \Omega_1(t_1) + \Omega_2(t_1) - \Omega_3(t_1)
\end{equation}
where
\begin{align*}
\Omega_1(t_1) & = \int_0^\tau K(v) \, d\alpha_3(v) \\[2mm]
\Omega_2(t_1) & = \int_0^\tau \left\{ -S_2(v-) \int_{[0,v)} \frac{S_{2-}}{S_2} 
\cY_2^{-1} [d\alpha_1 - \alpha_2 \cY_2^{-1} \, d\cN_2 ] \right\} \cY_2(v)^{-1} \, dQ(v;\cdot) \\[2mm] 
\Omega_3(t_1) & = \int_0^\tau \cY_2(v)^{-1} K(v) \, d\alpha_2(v;\cdot) 
\end{align*}
Next, define $Z_n = \sqrt{n} \, (\bar{N}_{2n}-\cN_2,\bar{Y}_{2n}-\cY_2,Q_n-Q)$.
By standard empirical process theory, $Z_n$ converges weakly to a mean-zero Gaussian process $Z$
with the same covariance structure as that of $Z_n$. Therefore, by the functional delta method
(Theorem 20.8 of \cite{vdv}), $\sqrt{n} \, (\widehat{G}_1(t_1)-G_1(t_1))$ converges weakly
to a Gaussian process and we can write
\begin{align*}
\widehat{G}_1(t_1) - G_1(t_1) & = \cM(\bar{N}_2,\bar{Y}_2,Q_n) - \cM(\cN_2,\cY_2,Q) \\
& = \cM^\pr(\bar{N}_{2n}-\cN_2,\bar{Y}_{2n}-\cY_2,Q_n-Q) + o_P(n^{-1/2}) \\
& = \Omega_1^*(t_1) + \Omega_2^*(t_1) - \Omega_3^*(t_1) + o_P(n^{-1/2})
\end{align*}
with
\begin{align*}
\Omega_1^*(t_1) & = \int_0^\tau K(v) \, d(Q_n(v;t_1) - Q(v;t_1)) \\[2mm]	
\Omega_2^*(t_1) & = \int_0^\tau \left\{ -S_2(v-) \int_{[0,v)} \frac{S_{2-}}{S_2} 
\cY_2^{-1} [d(\bar{N}_{2n}-\cN_2) - (\bar{Y}_{2n}-\cY_2) \cY_2^{-1} \, d\cN_2 ] \right\} \cY_2(v)^{-1} \, dQ(v;t_1) \\[2mm] 
 & = \int_0^\tau \left\{ -S_2(v-) \int_{[0,v)} \frac{S_{2-}}{S_2} 
\cY_2^{-1} [d(\bar{N}_{2n}-\cN_2) - \cY_2^{-1}\bar{Y}_{2n} \, d\cN_2 + d\cN_2 ] \right\} \cY_2(v)^{-1} \, dQ(v;t_1) \\[2mm] 
 & = \int_0^\tau \left\{ -S_2(v-) \int_{[0,v)} \frac{S_{2-}}{S_2} 
\cY_2^{-1} [d\bar{N}_{2n} - \cY_2^{-1}\bar{Y}_{2n} \, d\cN_2 ] \right\} \cY_2(v)^{-1} \, dQ(v;t_1) \\[2mm]
\Omega_3^*(t_1) & = \int_0^\tau \cY_2(v)^{-1} K(v) (\bar{Y}_{2n}(v)-\cY_2(v))\, dQ(v;t_1)
\end{align*}
The term $\Omega_1^*(t_1)$ corresponds to the asymptotic behaviour of (\ref{gs}).
The term $\Omega_2^*(t_1)$ is the contribution due to estimation of $S_2$, and the
term $\Omega_3^*(t_1)$ is the contribution due to estimation of $\cY_2$.

We can further write
\begin{equation}
\widehat{G}_1(t_1) - G_1(t_1) = \frac{1}{n} \sumi \Psi_i(t_1) + o_P(n^{-1/2})	
\end{equation}
where
\begin{equation}
\Psi_i(t_1) = \Omega_{1i}^*(t_1) + \Omega_{2i}^*(t_1) - \Omega_{3i}^*(t_1)
\end{equation}
with
\begin{align*}
\Omega_{1i}^*(t_1) & = \delta_{1i} \delta_{2i} I(V_{1i} \leq t_1) K(V_{2i}) - G_1(t_1) \\[2mm]	 	
\Omega_{2i}^*(t_1) & = \int_0^\tau \left\{ -S_2(v-) \int_{[0,v)} \frac{S_{2-}}{S_2} 
\cY_2^{-1} [dN_{2i} - \cY_2^{-1}Y_{2i} \, d\cN_2 ] \right\} \cY_2(v)^{-1} \, dQ(v;t_1) \\[2mm]
& = -\int_0^\tau \left\{
\cY_2(V_{2i})^{-1} \delta_{2i} I(V_{2i}< v) - \int_{R_i}^{V_{2i} \wedge v-} 
\left( \frac{S_{2}(s-)}{S_2(s)} \right) \cY_2(s)^{-2} d\cN_2(s) 
\right\} \\
& \hspace*{20mm} S_2(v-) \cY_2(v)^{-1} \, dQ(v;t_1) \\
\Omega_{3i}^*(t_1) & = \int_0^\tau \cY_2(v)^{-1} K(v) (Y_{2i}(v)-\cY_2(v))\, dQ(v;t_1)
\end{align*}
We have now exhibited $\widehat{G}_1(t_1) - G_1(t_1)$ as a sum of i.i.d.\ mean-zero terms plus
a negligible remainder.
We can estimate $\Var(\sqrt{n} \, \{\widehat{G}_1(t_1) - G_1(t_1) \})$ using the empirical estimator
\begin{equation}
s^2(t_1) = \widehat{\Var}(\sqrt{n} \, \{\widehat{G}_1(t_1) - G_1(t_1) \}) = 
\frac{1}{n} \sumi \widehat{\Psi}_i(t_1)^2
\end{equation}
where
$$
\widehat{\Psi}_i(t_1) = \widehat{\Omega}_{1i}^*(t_1) + \widehat{\Omega}_{2i}^* (t_1)- \widehat{\Omega}_{3i}^*(t_1)
$$
with
\begin{align*}
\widehat{\Omega}_{1i}^*(t_1) & = \delta_{1i} \delta_{2i} I(V_{1i} \leq t_1) \widehat{K}(V_{2i}) - \widehat{G}_1(t_1) \\
\widehat{\Omega}_{2i}^*(t_1) & = 
- \frac{1}{n} \sum_{j=1}^n \delta_{1j} \delta_{2j} I(V_{1j} \leq t_1) \bar{Y}_{2n}(V_{2j})^{-1} \widehat{S}_2(V_{2j}-) \\
& \hspace*{15mm} \times \left\{ 
\bar{Y}_{2n}(V_{2i})^{-1} \delta_{2i} I(V_{2i}< V_{2j}) - \int_{R_i}^{V_{2i} \wedge V_{2j}-} 
\left( \frac{\widehat{S}_{2}(s-)}{\widehat{S}_2(s)} \right)
\bar{Y}_{2n}(s)^{-2} d\bar{N}_{2n}(s)  \right\} \\
\widehat{\Omega}_{3i}^*(t_1) & = \frac{1}{n} \sum_{j=1}^n \delta_{1j} \delta_{2j} I(V_{1j} \leq t_1) \bar{Y}_{2n}(V_{2j})^{-1} 
\widehat{K}(V_{2j})(Y_{2i}(V_{2j})-\bar{Y}_{2n}(V_{2j})) \, .
\end{align*}



\section*{Representation of the Aalen-Johansen Estimator}

The Aalen-Johansen estimator can be represented in a similar way.
Recall the definition $N_{1i}(u) = \delta_{1i} I(V_{1i} \leq u)$.
Let us define the following additional notation: 

\vspace*{6pt}

\noindent
$n_0$ = number of subjects alive and free of disease at the time of recruitment  \\[1mm]
$\cA$ = set of indices of subjects alive and free of disease at the time of recruitment \\[1mm]
$S^*$ = survival function for time to first transition (to either diseased or dead), conditional on being alive and
free of disease at the time of recruitment \\[1mm]
$\hat{S}^*$ = the corresponding Kaplan-Meier estimate \\[1mm] 
$T_{3i} = \min(T_{1i},T_{2i})$ \\[1mm]
$V_{3i} = \min(T_{3i}, C_i)$ \\[1mm]
$\delta_i^* = I(T_{3i} \leq C_i)$ \\[1mm]
$\xi_i = I(i \in \cA) = I(T_{3i} \geq R_i)$ \\[1mm]
$\pi = P(T_{3i} \geq R_i)$ \\[1mm]
$\hat{\pi} = n_0/n$ \\[1mm]
$Y_{1i}^\circ(t) = \xi_i I(R_i \leq t)I(V_{3i} \geq t)$ \\[1mm]
$N_{1i}^\circ(t) = \xi_i \delta_{1i}I(V_{1i} \leq t)$ \\[1mm]
$\cY_1^\circ(t) = E[Y_{1i}^\circ(t)]$ \\[1mm]
$\cN_1^\circ(t) = E[N_{1i}^\circ(t)]$ \\[1mm]
$\bar{Y}_1^\circ(t) = n^{-1} \sum_{i=1}^n Y_{1i}^\circ(t)$ \\[1mm]
$\bar{N}_1^\circ(t) = n^{-1} \sum_{i=1}^n N_{1i}^\circ(t)$ \\[2mm]
$\cY_1^*(t) = \pi^{-1} \cY^\circ(t)$ \\[1mm]
$\bar{Y}_1^*(t) = \hat{\pi}^{-1} \bar{Y}_1^\circ(t)
= n_0^{-1} \sum_{i \in \cA} Y_{1i}$(t) \\[1mm]
$K^\dagger(t) = S^*(t-)/\cY_1^\circ(t)$ \\[1mm]
$\hat{K}^\dagger(t) = \hat{S}^*(t-)/\bar{Y}_1^\circ(t)$ \\

\vspace*{6pt}
The Aalen-Johansen estimator can be written as

\allowdisplaybreaks
\begin{align*}
\hat{G}_1^\pss{AJ}(t_1) & =  \int_0^{t_1} \hat{S}^*(u-) \,
\frac{\sumi dN_{1i}^\circ(t)}{\sumi Y_{1i}^\circ(t)} \\
& = \int_0^{t_1}  \hat{K}^\dagger(u) d\bar{N}_1^\circ(u) \\
& = \int_0^{t_1} K^\dagger(u) d\bar{N}_1^\circ(u) +  \int_0^{t_1} (\hat{K}^\dagger(u) - K^\dagger(u)) d\bar{N}_1(u) \\
& = \int_0^{t_1} K^\dagger(u) d\bar{N}_1^\circ(u) +  \int_0^{t_1} \bar{Y}_1^\circ(u)^{-1} 
\left( \hat{S}^*(u-) - K^\dagger(u) \bar{Y}_1^\circ(u) \right) d\bar{N}_1^\circ(u) \\
& = \int_0^{t_1} K^\dagger(u) d\bar{N}_1^\circ(u) +  \int_0^{t_1} {\cY}_1^\circ(u)^{-1} 
\left( \hat{S}^*(u-) - K^\dagger(u) \bar{Y}_1^\circ(u) \right) d\cN_1^\circ(u) + o_P(n^{-1/2}) \\
& = \int_0^{t_1} K^\dagger(u) d\bar{N}_1^\circ(u) \\
& \hspace*{20pt} +  \int_0^{t_1} {\cY}_1^\circ(u)^{-1} 
\left([ \hat{S}^*(u-)-S^*(u-)] - K^\dagger(u) [\bar{Y}_1^\circ(u)-\cY_1^\circ(u)] \right) d\cN_1^\circ(u) + o_P(n^{-1/2}) \\ 
& = G_1(t_1) + \frac{1}{n} \sumi  \Psi_i^\pss{AJ}(t_1) + o_P(n^{-1/2}) 
\end{align*}
where for $i \in \cA$ we define
\begin{align}
\Psi_i^\pss{AJ}(t_1) & = [\delta_{1i} I(V_{i1} \leq t_1) K^\dagger(V_{1i})  - G_1(t_1)] \nonumber \\
& \hspace*{10mm} - \frac{1}{\pi} \int_0^{t_1} K^\dagger(u)
\left\{ 
\frac{\delta_i^*}{\cY^*(V_{3i})}I(V_{3i} < u) \right. \nonumber \\
& \hspace*{20mm} - \left. \int_{R_i}^{V_{3i} \wedge u-} \left( \frac{S^*(s-)}{S^*(s)} \right) \cY^*(s)^{-2} d\cN^*(s) \right\} d\cN_1^\circ(u) \nonumber \\
& \hspace*{20mm}  - \int_{0}^{t_1} {{\cY}_1^\circ(u)}^{-1} K^\dagger(u) (Y_{1i}^\circ(u)- \cY_1^\circ(u)) \, d\cN_1^\circ(u)
\end{align}
and for $i \notin \cA$ we define $\Psi_i^\pss{AJ}(t_1)=0$.
An estimate $\hat{\Psi}_i^\pss{AJ}(t_1)$ of $\Psi_i^\pss{AJ}(t_1)$ for $i \in \cA$
can be obtained by replacing $S^*$ by $\hat{S}^*$, $\cN_1^\circ$ by $\bar{N}_1^\circ$, $\cN^*$ by $\bar{N}^*$,
$\cY_1^\circ$ by $\bar{Y}_1^\circ$, $\cY^*$ by $\bar{Y}^*$,  and $G_1$ by $\hat{G}_1$. 
This yields 
$$
\hat{\Psi}_i^\pss{AJ}(t_1) = \hat{\Omega}_{1i}^\pss{AJ}(t_1) + \hat{\Omega}_{2i}^\pss{AJ} (t_1)
- \hat{\Omega}_{3i}^\pss{AJ}(t_1)
$$
with
\begin{align*}
\hat{\Omega}_{1i}^\pss{AJ}(t_1) & = \delta_{1i} I(V_{1i} \leq t_1) \hat{K}^\dagger(V_{1i}) - \hat{G}_1(t_1) \\
\hat{\Omega}_{2i}^\pss{AJ}(t_1) & = 
- \frac{1}{\hat{\pi}} \sumj \xi_j \delta_{1j} I(V_{1j} \leq t_1) \hat{K}^\dagger(V_{1j})  \\
& \hspace*{30pt} \times \left\{ 
\bar{Y}^*(V_{3i})^{-1} \delta_i^* I(V_{3i}< V_{1j}) - \int_{R_i}^{V_{3i} \wedge V_{1j}-} 
\left( \frac{\hat{S}^*(s-)}{\hat{S}^*(s)} \right)
\bar{Y}^*(s)^{-2} d\bar{N}^*(s)  \right\} \\
\hat{\Omega}_{3i}^\pss{AJ}(t_1) & = \frac{1}{n} \sumj \xi_j \delta_{1j} I(V_{1j} \leq t_1) 
\bar{Y}^\circ(V_{1j})^{-1} \hat{K}^\dagger(V_{1j})
(Y_{1i}^\circ(V_{1j})-\bar{Y}_1^\circ(V_{1j}))
\end{align*}
We can then construct pointwise confidence intervals and a simultaneous confidence band based on $\hat{G}_1^\pss{AJ}$ 
using procedures similar to those used for $\hat{G}_1$.

\end{document}